\magnification1200

\rightline{KCL-MTH-10-06}
%\rightline{hep-th/yymmnnn}

\vskip .5cm
\centerline
{\bf ÊÊGeneralised space-time and duality }
\vskip 1cm
\centerline{Peter West}
\centerline{Department of Mathematics}
\centerline{King's College, London WC2R 2LS, UK}

%\vskip 0.5cm
%\centerline{and}
%\vskip 0.5cm
%\centerline{??,}
%\centerline {??,}
%\centerline{??}
%\leftline{\sl Abstract}
\vskip .2cm
\noindent
In this paper we consider the previously proposed generalised
space-time and investigate the structure of the field theory upon which
it is Êbased. In particular, we derive a SO(D,D) formulation of the
bosonic string as a non-linear realisation at lowest levels of
$E_{11}\otimes_s l_1$ where $l_1$ is the first fundamental
representation. We give Êa Hamiltonian formulation of this theory
and carry out its quantisation. We argue that the choice of
representation of the quantum theory breaks the manifest SO(D,D)
symmetry
but that the symmetry is manifest in a non-commutative field theory. ÊWe
discuss the implications for
the conjectured $E_{11}$ symmetry and the role of the $l_1$
representation.
\vskip .5cm

\vfill
\eject

\medskip
{\bf 1. Introduction}
\medskip
When it was first conjectured that Êthe maximal
supergravities in any dimensions could be extended to possess an
$E_{11}$ symmetry [1] it was realised that some modification of
space-time would be required rather than the ad hoc introduction of the
translation generators as was done in the early papers [1,2]. It was
subsequently proposed [3] that one introduce the generators
transforming in Êthe fundamental
representation $l_1$ Êof $E_{11}$; more precisely owe should
take the non-linear realisation of the semi-direct product of $E_{11}$
and the
$l_1$ representation, i.e. $E_{11}\otimes_s l_1$ [2]. At lowest levels
the $l_1$ Êmultiplet in eleven dimensions Êbegins with the space-time
translation generators
$P_a$ , Êthen a two form $Z^{a_1a_2}$, Êfive form generator
$Z^{a_1\dots a_5}$ and a generator $Z^{a_1\dots a_7,b}$ together with an
infinite number of other generators. ÊIn this approach the
fields would depend on all the coordinates introduced in this non-linear
realisation ÊÊthat is
$x^a, x_{a_1a_2}, x_{a_1\ldots a_5}, x_{a_1\dots a_7,b},\ldots$ [3].
\par
The simplest application of this idea is to consider the reduction
on a circle, Êthat is the IIA theory, Êand restrict the $E_{11}$ algebra
to the
$D_{10}$ subalgebra found by deleting node ten. The Dynkin diagram of
$E_{11}$ is given in Êfig 1. In this case,
at lowest level, Êone has the coordinates $x^a, y_a, a=1,\ldots 10$,
arising as the dimensional reductions of
$x^a$ and
$x_{a11}$, Êbelonging to the vector representation Êof
$D_{10}$ and the field content $h_a{}^b, B_{a_1a_2}, \phi$ which should
now depend on Êthese coordinates which belong to the vector
representation of SO(D,D) [4]. Although the non-linear realisation that
arises is relatively straightforward to work out, one would be Êleft
with the problem of how to recover the usual theory, that is the
massless
NS-NS sector of the superstring. The introduction of the coordinates
$x^a$ and
$y_a$ Êhas a long history in the context of $T$ duality in string
theory;
two of the earliest papers being Ê[5,6]
\par
The content of the $l_1$ representation can be found [7] by considering
the algebra whose Dynkin diagram Êis that for
$E_{11}$ but with one node added to the node labeled one of the $E_{11}$
Dynkin diagram by a single line and taking, with respect to the new
node,
only level one generators in the enlarged algebra. ÊIn eleven
dimensions we decompose the
$l_1$ representation in terms of representations of
$A_{10}$ and we find that at lowest levels the
generators [7]
$$
P_{\hat a}; Z^{\hat a_1\hat a_2}; Z^{\hat a_1
\ldots\hat Êa_5}; Z^{\hat a_1\ldots \hat Êa_7,b}, ÊZ^{\hat
ÊÊa_1\ldots \hat Êa_8}; Z^{\hat Êa_1\ldots \hat Êa_8, \hat b_1\hat
b_2\hat b_3}, Z^{\hat Êa_1\ldots \hat
ÊÊa_9,(\hat b\hat c)}, Z^{\hat Êa_1\ldots \hat a_9,\hat b_1\hat b_2},
ÊÊ$$
ÊÊ$$
ÊÊZ^{\hat Êa_1\ldots \hat a_{10},\hat b}, Z^{\hat Êa_1\ldots \hat a_
{11}}; Z^{\hat Êa_1\ldots \hat a_9,\hat
ÊÊb_1\ldots \hat b_4,\hat c},
Z^{\hat Êa_1\ldots \hat a_8,\hat b_1\ldots \hat b_6}, Z^{\hat Êa_1\ldots
\hat
ÊÊa_9,\hat b_1\ldots \hat b_5},\dots
ÊÊ$$
where $ \hat a=1,\ldots ,11$.

By deleting node labeled
$d$ in the Dynkin diagram of the enlarged algebra one can find the
content
of the
$l_1$ representation appropriate to the
$d$ dimensional theory, that is decomposed in terms of
representations of
$E_{11-d}\otimes GL(d)$. The results [4,8,9] are given in table one.
Indeed Êpage 13 of the second of these papers Êcontains the pont
particle multiplet in all dimensions Êthree and above. One can also find
this result at low levels by Êsimply carrying Êout the dimensional
reduction by hand on the eleven dimensional generators contained in the
above equation. In the non-linear realisation of $E_{11}\otimes_s
l_1$ in $d$ dimensions one would introduce coordinates
corresponding to the charges in the table.
\par
The first few generators of the $l_1$ representation in eleven
dimensions are the charges associated with the Êpoint particle, the
two brane and the five brane and it has been proposed that the $l_1$
representations contains all the brane charges [4,7] in eleven and lower
dimensions. One piece of evidence supporting this conjecture is that for
every $A_{10}$ representation in Êthe adjoint representation of
$E_{11}$, that is gauge field in the non-linear realisation, Êwe can
find
an Ê$A_{10}$ representation in the $l_1$ multiplet that has the
correct space-time index structure to be interpreted as the charge
corresponding to the current of the brane to which the gauge field
couples. Put another way the existence of ÊWess-Zumino term in the
dynamics of branes implies Êa pairing between gauge fields and currents
and this implies a correspondence between the representations in the
adjoint and
$l_1$ representations of $E_{11}$ that holds [7].
\par
By taking a particular charge found by dimensional reduction Êand
applying $U$ duality in $d$ dimensions some charge multiplets for point
particles and some other branes Êhave been previously found [10-13].
For example, Êfor the point particle we can take the charge of one of
the
the Kaluza Klein particles, that is ${n\over R_i}$ where
$R_i$ is one of the radii of the torus used in the ÊÊdimensional
reduction.
The results
predicted by
$E_{11}$ agree with these mutliplets, ÊÊthat is Êthe first two
columns of
table 1 agree with the point particle and string mutliplets found
earlier. ÊWhile it is obvious that the decomposition of the
$l_1$ multiplet would lead to multiplets of
$E_{11-d}\otimes GL(d)$ it did not have to lead to the correct
representations. Put another way the charge representations in $d$
dimensions Êdid not have to assemble into a single representation of
$E_{11}$. Thus there is considerable evidence that the
$l_1$ multiplet does contain all the brane charges. The table of
figure 1
predicts Êquite a few other charge mutliplets and it would be good to
understand their role in string theory
\par
One can interpret the dependence on the generalised coordinates as
encoding the measurement of events by all the different branes using the
space-time that they see, that is $x^a$ corresponds to a point particle,
$x^a$ and $ x_{a_1a_2}$ by a two brane etc. Thus
encoding all the coordinates of the $l_1$ multiplet allows one to
consider all the possible ways of measuring space-time using all the
different probes in a way that Êreflects
the underlying symmetry that is $E_{11}$ and in particular, in lower
dimensions, $U$ duality. The approach of introducing a generalised
space-time was used to construct the field strengths of all maximal
gauged supergravities in five dimensions [14]. In particular this paper
introduced the coordinates Ê$ x_N,\
x_a^N,
\ x_{a_1a_2\alpha}, x_{a_1 a_2} ,
\ x_{a_1a_2a_3}^{NM} , \ x_{a_1a_2a_3 N} Ê,\ldots$ in addition to those
of the usual five dimensional space-time, although the dependence on
these coordinates was of a rather specific form.
However, it is far from clear how to recover our usual supergravity
theories in the general situation and it is likely that one is over
counting in some way by introducing all the coordinates. It is the
purpose of this paper to try to shed some light on this dilemma.
\par
Many of the features suggested in references [3,7,4], Êand discussed
above, Êhave appeared in subsequent works on generalised geometry. It
would be invidious to reference these papers here.

\par
It has also been proposed [8,15] to use a non-linear realisation based
on
$E_{11}\otimes_s l_1$ to give a description of brane dynamics. Although
the algebra is the same as that used for the supergravity
theories the dependence on the fields Êand the choice of local
subalgebra is different. ÊWe carry out this non-linear realisation in
section 2 taking the IIA perspective and only the lowest level as
described above, that is the SO(D,D) algebra Êand Êthe brane coordinates
$x^a, y_a, a=1,\ldots , 10$. We Êarrive at a formulation of string
theory
that has SO(D,D) symmetry and constructed from Êthe coordinates $x^a,
y_a$. It is in fact a formulation found long ago [5].
\par
The quantisation of the Êdynamics of the usual string based on the
coordinate $x^\mu$, that is the Nambu action, leads to a quantised
theory
[16] that contains the bosonic string. In this paper we will
quantise Êthe SO(D,D) string just mentioned. ÊÊIn
section 3 we find ÊÊits ÊHamiltonian formulation and in section
4 we quantise this theory to find that the $x^a$ and $y_a$
coordinates do
not commute. In Êthe Êquantised theory one can work with just
$x^a$ but then the SO(D,D) symmetry is not manifest. To maintain
manifest SO(D,D) symmetry we must work with both
$x^a$ and
$y_a$, but then one is dealing with a non-commutative field theory. ÊIn
section five we discuss the implications of this work for the $E_{11}$
conjecture and the role of the $l_1$ representation.

\medskip
{\bf 2. SO(D,D) symmetric string as a non-linear realisation}
\medskip
Let us begin by briefly summarising the two methods of carrying out a
non-linear realisation for an internal group, that is where the
space-time coordinates are inert under the group involved [17]. In
particular we consider the non-linear realisation of a group
$G$ with local sub-algebra
$H$ with a group element $g\in G$ which depends on a set of
parameters which in turn depend on the ÊÊspace-time coordinates. ÊThus
the parameters in the group element become the fields of the theory. The
group element
$g$ is subject to the two transformation
$g\to g_0 g$ Êand $g\to Êg h$ where
$g_0\in G$ Êand $h\in H$, but while $g_0$ is a rigid transformation and
so is independent of the coordinates of space-time, the $h$
transformation is a local transformation and so does depend on the
space-time coordinates. To construct the non-linear realisation we must
find Êsome dynamics which is invariant under the above two
transformations. ÊÊLet us assume that
$G$ is
a Kac-Moody algebra and that the local sub-algebra
$H$ is the one invariant under the Cartan
involution $I_c$ which acts on Êthe Chevalley generators $E_a$, $F_a$
and
$H_a$ as
$$
I_c(E_a)=-F_a,\ I_c(F_a)=-E_a,\ I_c(H_a)=-H_a,\
\eqno(2.1)$$
Another useful operator, denoted by
$I$, ÊÊacts on group elements as
$I(g)=I_c(g^{-1})$ and on element $A$ of the algebra by $I(A)=I_c(-A)$.
We note that this operator Êalso squares to the identity operation
and that
$I(A B)=I(B)I(A)$ for any two elements $A$ and $B$ of the Lie algebra.
Since by definition $h$ satisfies $I_c(h)=h$ it follows that
$I(h)=h^{-1}$. Using the local $H$ symmetry we may set to zero the $H$
part of $g$, leaving Êit to be a member of the Borel sub-algebra of
$G$. Having done this one must then carry out a local $h$
transformations for a generic Ê$g_0$ transformations to preserve the
choice of coset representative.
\par
Since there are two distinct symmetries that must be taken into
account, that is the Êabove rigid and local symmetries, there two are
ways to proceed. One can first find objects which are invariant
under
$g_0$ transformations and then solve the invariance with respect to $h$
transformations. This is achieved by considering the Cartan forms
${\cal V}=g^{-1}d g$ which are indeed invariant under the former
transformations and transform under local $H$ transformations as
${\cal V}\to h^{-1}{\cal V} h+h^{-1}d h$. To construct the dynamics we
exploit the properties of the Cartan involution and consider the objects
$$
{\cal U}={\cal V}+I({\cal V})=d\xi^\alpha Ê{\cal U}_\alpha, \ \ w={1
\over
2}({\cal V}-I({\cal V}))=d\xi^\alpha Ê{w}_\alpha
\eqno(2.2)$$
which transform as
$$
{\cal U}\to h^{-1}{\cal U} h,\ w\to ÊÊh^{-1}w h+ Êh^{-1}d h
\eqno(2.3)$$
We note that $D_\alpha {\cal U}_\beta\equiv Ê\partial_\alpha {\cal
U}_\beta+[{w}_\alpha,{\cal U}_\beta]$ transforms covariantly
; $D_\alpha {\cal U}_\beta\to h^{-1}D_\alpha {\cal U}_\beta h$.
An invariant Êis given by
$$
Tr ({\cal U}_\alpha {\cal U}_\beta)
\eqno(2.4)$$
Integrating this expression over space-time and contracting the space-
time
indices we can take this to be the action.
\par
Alternatively, one can first find objects
invariant under the local $H$ transformations. Such an object is
$M\equiv gI(g)$ which is invariant Êunder local $h$ as $I(h)=h^{-1}$ and
transforms under the rigid transformations as
$M\to g_0 M I(g_0)$. Clearly,
$$
tr (M^{-1}\partial_\alpha M M^{-1}\partial_\beta ÊÊM )
\eqno(2.5)$$
is invariant under the local and rigid transformations.
In fact the two invariants of equations (2.4) and (2.5) are the same
up to
a constant of proportionality. Such Êconstructions Êhas proved
particularly useful in formulating the symmetries of the scalars in the
dimensionally reduced maximal supergravity theories.
\par
We can also consider a non-linear realisation in which the generators of
space-time belong to the group $G$. Such is the case for
gravity and supergravity. In particular the group used in Êsuch
non-linear realisations is extended to include generators that belong
to a
realisation $l$ of
$G$ and consider the semi-direct product group formed from $l$ and
$G$, denoted $G\otimes_s l$. In terms of the algebra, if the generator
$A$ is in the Lie algebra $G$ and Êthe generators in the $l$
representation are denoted by $Z_s$ then we adopt the commutator
$[Z_s , A]=D(A)_s{}^r Z_r$ where $D(A)$ is the matrix representation
of $G$
of the $l_1$ representation. We could take the generators Ê$Z_s$ to have
non-trivial commutators amongst themselves provided this is
consistent with the Jacobi identities, but here we will take them to
commute. The group element can be written in the form
$$
g=g_l g_B
\eqno(2.6)$$
where $g_l$ is Êgenerated by the $Z_r$ the coefficients of which are the
the Êspace-time coordinates and the $g_B$ belongs to $G$ and the
coefficients of the generators of $G$ are the fields of the theory which
are taken to depend on the coordinates of the generalised Êspace-
time. As
above one has the two transformations, one rigid and one local. The
local
sub-algebra being the Cartan sub-algebra of $G$. For examples of how
this
method proceeds see references [2].
\par
Now let us consider the non-linear realisation appropriate to a brane
moving in a background, Êsuch as gravity and supergravity [2]. The
background fields belong to a non-linear realisations as described just
above, that is the arise as a non-linear realisation of a Lie Êalgebra
$G\otimes_s l$.
The local sub-algebra is chosen to be a sub-algebra $H_a$ of the
Cartan involution invariant sub-algebra $H$ of $G$. This corresponds to
the breaking of some of the background symmetries by the presence of the
brane. Different Êchoices of this local sub-algebra give rise to
different branes. The group element is of the form
$$
g=g_l g_Bg_a
\eqno(2.7)$$
where $g_l$ belongs to the Abelian ÊÊgroup generated by $l$ i.e. by
the
$Z_r$, Êand it contains the space-time coordinates; Ê$g_b$ is a group
element of
$G$ which has its
$H$ part removed, so belongs to
the Borel sub-algebra of
$G$ and it contains the background fields. Finally
$g_a$ is a group element of
$H$ and it contains more fields associated with the breaking of the
symmetries of $G$ Êby the brane. We may remove the Êpart of
$g_a$ that belongs to
$H_a$ using the local sub-algebra.
\par
The brane coordinates, Êwhich appear in
$g_l$, depend on the parameters Êthat describe the Êworld
volume of the brane. The background fields, that appear in $g_B$,
depend on the brane coordinates and $g_a$ contains further fields,
denoted
$\phi$ Êthat also depend on the parameters $\xi$ of the brane world
volume. ÊÊAs $l$ arises as a representation of $G$, ÊÊunder a $g_0$
transformation
$g_l\to g_0 g_l (g_0)^{-1}$ and so Êthey transform linearly under $G$.
\par
Of course the group $G\otimes_s l$ is Ênot a Kac-Moody algebra, or in
the
finite case a Êsemi-simple Lie algebra,
nonetheless one can extend the notion of Cartan involution and Êthe
involution $I$ Êto act on Êthe generators $Z_r$ to give new generators
$\tilde Z_r$ and then use this extended involution to construct
invariant dynamics. One can do this by trial and error until one
finds an
involution of the algebra
$G\otimes_s l$ which reduces to the previous involution when
restricted to $G$.
However, when Êthe representation
$l$ is one of the fundamental representation of $G$ we can enlarge the
algebra by adding an extra Ênode to the Dynkin diagram of $G$
attached to
the node associated with the fundamental representation. The generators
of the $G$ are the level zero generators of the new algebra and, as
explained in [7], Êthe generators of the $l$ representation are those of
level one. The action of the ÊCartan involution takes these generators
into those at level minus one. Thus one can embed Êthe generators of
$G\otimes_s l$ into a ÊKac-Moody algebra whose Dynkin diagram is the
enlarged Dynkin diagram just discussed and then we can Êuse the
Cartan involutions Êfor any Kac-Moody algebra.
\par
The situation is most readily understood by considering the simplest
example; the bosonic p brane in space-time dimension $D$ coupled to
gravity. For this case $G=GL(D)$, $H=SO(1,p)\otimes SO(D-p-1)$
the $l$ representation is
the fundamental representation associated with first node of the Dynkin
diagram of $A_{D-1}$, which are just the usual translations $P_a$.
The group
element has the form
$$
g=e^{x^a(\xi) P_a} e^{K^a{}_b h_a{}^b(x^c)}e^{\phi(\xi)\cdot J}
\eqno(2.8)$$
where the three group elements correspond to the decomposition
of equation (2.7) and so ${\phi\cdot J}$ contains terms that only belong
to $SO(1,D)$ moded out by $SO(1,p)\otimes SO(D-p-1)$ and $K^a{}_b$ is
symmetric once one of its indices is lowered. Using the suitable
generalisation of the Êfirst method discussed above for internal
symmetries Êto construct the non-linear realisation, Êthat is worked
with
${\cal V}$, one finds Êcovariant constraints that
express the fields
$\phi$ in terms of the derivatives of the coordinates
$x^a$. ÊÊand construct Êan invariant actions out of these remaining
fields. As such one
ends up with an invariant action that just contains the the field $x^a$
and the background metric
$g_{mn}=e_m{}^a \eta_{ab} e_n{}^b$ where $e_m{}^a =(e^h)_m{}^a$. It is
just the Êvolume swept [2].
\par
The above example is thought to illustrate Êthe general situation,
one can
find covariant constraints such that the fields
$\phi$ are expressed in terms of derivatives of the brane coordinates.
However, there is at present no systematic way of choosing $H_a$ or of
finding Êthese constraints. ÊAs such we will use the analogue of the
second of the above methods, namely that based on
$M=gI(g)$. This has a substantial advantage in that the $g_a$ part of
the
group element drops out of $M$ and so one does not need to know the
$H_a$ or need to find the covariant constraints that express $\phi$ in
terms of the derivatives of $X^a$.
\par
We illustrate this for a bosonic p brane Êcoupled to gravity. The
algebra of GL(D) is given by
$$
[K^a{}_b,K^c{}_d]=\delta _b^c K^a{}_d - \delta _d^a K^c{}_b,
\eqno(2.9)$$
and their relations with the commuting translations $P_a$ are
$$
[K^c{}_b, P_a]=-\delta_a^c P_b
\eqno(2.10)$$
\par
The Cartan involution acts as $I_c(K^a{}_b)=-K^b{}_a$ and we identify
the Cartan involution invariant sub-algebra of SL(D) to be generated by
$J_{ab}=K_{ab}-K_{ba}$ which is the algebra SO(D). Acting on the
translations we find
$I_c(P_a)=-\tilde P^a$ where
$\tilde P^a$ is a new generator. Since the algebra is invariant under
the
Cartan involution this implies the commutation relation
$$
[K^b{}_a, \tilde P_c]=\delta_a^c \tilde P^b
\eqno(2.11)$$
The involution $I$ acts as Ê$I(K^a{}_b)=K^b{}_a$ and $I(P_a)=\tilde
P^a$ and so its action on Êthe group element of equation (2.8) is given
by
$$
I(g)=e^{-\phi(\xi)\cdot J}e^{K^a{}_b h^b{}_a(x^c)}e^{x^a(\xi)\tilde
P_a}
\eqno(2.12)$$
The quantity $M=gI(g)$ transforms as $M\to g_0 M I(g_0)$ Êand one finds
that ÊÊin a constant background
$$
M^{-1}\partial_\alpha M= \partial_\alpha x^a\tilde P^a
+ \partial_\alpha x^n g_{nm}
\eqno(2.13)$$
where
$$
g_{nm}=e_n{}^a e_{am},\ \ {\rm and }\ \ e_n{}^a=(e^h)_n{}^a
\eqno(2.14)$$
The invariant Ê$Tr (M^{-1}\partial_\alpha M M^{-1}\partial_\beta ÊÊM )
$ is easily evaluated
$$
\gamma_{\alpha\beta}=-{1\over 2}tr (M^{-1}\partial_\alpha M
M^{-1}\partial_\beta ÊÊM ) =\partial_\alpha x^n g_{nm}\partial_\beta x^m
\eqno(2.16)$$
In doing this we have used that $Tr (P_a\tilde P^b)=\delta _a^b$ and
$Tr (P_a P_b)=0=Tr (\tilde P^a\tilde P^b)$.
The reparametrisation invariant and rigid $G$ and local $H_a$ invariant
action is
$$
\int d^{p+1}\xi det(-\gamma_{\alpha\beta})
\eqno(2.16)$$
which is the well known action for the motion of a bosonic brane.
\par

\par
We now construct, the ISO(D,D) invariant string from a non-linear
realisation. As explained above this is just the non-linear
realisation of
$E_{11}\otimes_sl_1$, as seen from the IIA perspective and at the lowest
level. ÊTo this end we take
$G\otimes_s l$ to be the group ISO(D,D) whose Êcommutation relations are
given by
$$
[K^a{}_b,K^c{}_d]=\delta _b^c K^a{}_d - \delta _d^a K^c{}_b, Ê\ \
[K^a{}_b, R^{c d}]=\delta_b^c R^{ad}-\delta_b^d R^{ac},
[K^a{}_b, \tilde R_{c d}]=-\delta_c^a \tilde R_{bd}+\delta_d^a \tilde
R_{bc},
$$
$$
[R^{ab}, \tilde R_{c d}]=\delta_{[c}^{[a}
K^{b]}{}_{d]},\ \ \ [R^{ab}, R^{c d}]=0=[\tilde R_{ab}, \tilde R_{c d}]
\eqno(2.17)$$
$$
[K^c{}_b, P_a]=-\delta_a^c P_b,\ \ Ê[ ÊR^{ab}, P_c]=-{1\over 2}
(\delta^a_c Q^b-\delta^b_c Q^a),\ \ Ê[\tilde ÊR_{ab}, P_c]=0,
\eqno(2.18)$$
$$
[K^a{}_b, Q^c]=\delta_b^c Q^a,\ \ Ê[ \tilde R_{ab}, ÊQ^c]={1\over
2} (\delta_a^c P_b-\delta_b^c P_a),\ \ Ê[R^{ab},Q^c]=0,
\eqno(2.19)$$
The $l$ representation is the fundamental representation associated
with the node of SO(D,D) Dynkin diagram that is on the end of the long
tail. It corresponds to the generators
$P_a,Q^a$ and ISO(D,D) is indeed the corresponding
semi-direct product group. ÊThe Cartan involution acts on
the generators of ÊSO(D,D) as $I_c(K^a{}_b)=-K^b{}_a$ Êand
$I_c(R^{ab})=-\tilde R_{ab}$ and the corresponding invariant sub-algebra
is generated by $J_{ab}= K^a{}_b -K_b{}^a$ and $S_{ab}=2(R_{ab}-\tilde
R_{ab})$. It is just the algebra SO(D)$\otimes$SO(D).
Acting with the Cartan involution on the generators in the $l$
representation we find new generators $\tilde P^a,\tilde Q_a$ such that
$I_c(P_a)=-\tilde P^a,I_c(Q^a)=-\tilde Q_a$. We take them to commute
with $P_a,Q^a$ Êand their commutation relations with the generators of
SO(D,D) are given by
$$
[K^a{}_b, \tilde P^c]=\delta_b^c \tilde P^a,\ \ Ê[\tilde ÊR_{ab},
\tilde P^c]={1\over 2} (\delta_a^c \tilde Q_b-\delta_b^c \tilde Q_a),\ \
[R^{ab},\tilde P^c]=0,
\eqno(2.20)$$
$$
[K^a{}_b, \tilde Q^c]=-\delta_c^a \tilde Q_b,\ \ Ê[ R^{ab},
\tilde Q_c]=-{1\over 2} (\delta^a_c \tilde P^b-\delta_a^c \tilde P^b),
\ \
[\tilde R_{ab}, \tilde Q^c]=0,
\eqno(2.21)$$
\par
We note that the full algebra of equations (2.17) to (2.21) admits
another
automorphism $J$ which leaves the generators of SO(D,D) invariant but
acts as $J(P_a)=\tilde Q_a, J(Q^a)=\tilde P^a$.
\par
We can now construct the non-linear realisation for the string, as
discussed above. We take
$G=SO(D,D)$ with $H=SO(D)\otimes SO(D)$ and $H_a=SO(D-2)\otimes
SO(D-2)\otimes SO(1,1)\otimes SO(1,1)$. The corresponding group element
is given by
$$
g=e^{x^a(\xi) P_a+y_a (\xi) Q^a} e^{K^a{}_b h_a{}^b(x^c)}
e^{B_{bc}(x^c) R^{bc}}e^{\phi(\xi)\cdot J}
\eqno(2.22)$$
In fact this is not quite the complete Êlowest Êlevel $E_{11}\otimes_s
l_1$ realisation as we have discarded the dilaton and its associated
Abelian generator. Its inclusion would not significantly alter the
conclusions.
\par
It is straightforward to calculate, in a constant background, Êthe
locally
$H$ invariant object
$$
M^{-1}\partial_\alpha M=\partial_\alpha y_n \tilde Q^n+ \partial_\alpha
x^n \tilde P^n+\nabla_\alpha x_m P_m +\nabla_\alpha y^m Q^m
\eqno(2.23)$$
where
$$
\nabla_\alpha x_m=\partial_\alpha Êx^pg_{pm}-\partial_\alpha ÊÊy_nB^n
{}_m
-\partial_\alpha Êx^pB_{pq}B^q{}_m ,\quad
\nabla_\alpha y^m=\partial_\alpha Êy_p g^{pm}+\partial_\alpha
x^nB_n{}^m
\eqno(2.24)$$
\par
Under a rigid $g_0$ transformation $M^{-1}\partial_\alpha M\to
(I(g_0))^{-1}M^{-1}\partial_\alpha M I(g_0)$. We may write $g_0$
as $g_0=kl_0$ where $k\in SO(D,D)$ and $l_0$ is generated by $P_a,Q^a$.
As Êthe latter Êcommute with themselves and with $\tilde P_a,\tilde Q^a$
we find that Ê$M^{-1}\partial_\alpha M\to
(I(k))^{-1}M^{-1}\partial_\alpha M I(k)$. Now consider the above
automorphism $J$, as it leaves elements of SO(D,D) inert we find that
$J(M^{-1}\partial_\alpha M)$ transforms just like $M^{-1}\partial_\alpha
M$. Consequently, Êthe first order equations of motion
$$
\epsilon ^{\alpha\beta}J(M^{-1}\partial_\beta M)=\sqrt{-\gamma}
\gamma^{\alpha\beta} M^{-1}\partial_\beta M
\eqno(2.25)$$
are invariant under local and rigid transformations provided we can find
a $\gamma_{\alpha\beta}$ which is also invariant and transforms like a
metric under two dimensional reparmeterisations.
Leaving aside this one point for the moment Êthe above
equations of motion are given in terms of the Êfields by
$$
\epsilon ^{\alpha\beta}\partial_\beta y_n=\sqrt{-\gamma}
\gamma^{\alpha\beta}\nabla_\beta x_n,\ \ \
\eqno(2.26)$$
$$
\epsilon ^{\alpha\beta}\partial_\beta x^n=\sqrt{-\gamma}
\gamma^{\alpha\beta}\nabla_\beta y^n .
\eqno(2.27)$$
The other two equations contained in equation (2.25) are Êequivalent to
these two equations as is to be expected as the equation is invariant
under the action of $J$. ÊIn fact one requires only one of the above
equations as the first Êimplies the second. These equations were first
given in reference [5]. We note that the doubling of coordinates is
similar to the introduction of dual fields, such as for
electromagnetism; they Êboth allow the equation of motion to be
written in two ways and these together form a mulitplet under the
duality group.
\par
At first sight equations (2.26) and (2.27) provide a manifestly SO(D,D)
set of equations of motion for the string, Êhowever, this assumes one
can
find an expression for $\gamma_{\alpha\beta}$ that is manifestly SO(D,D)
invariant. We have at our disposal the invariant
$tr (M^{-1}\partial_\alpha M M^{-1}\partial_\beta ÊÊM )$ which, using
equation (2.23) evaluates to
$$
I^1_{\alpha\beta}\equiv tr (M^{-1}\partial_\alpha MM^{-1}\partial_\beta
M )
=\partial_\alpha x^n\nabla_\beta x_n+\partial_\alpha y_n\nabla_\beta y^n
+(\alpha\leftrightarrow \beta)
\eqno(2.28)$$
This would seem, at first sight, a good choice for
$\gamma_{\alpha\beta}$. ÊHowever,
multiply ÊÊequation Ê(2.26) by $\partial_\gamma x^n$ and ÊÊequation
(2.27) by $\partial_\gamma y_n$ and adding we find, multiplying by
$\epsilon_{\delta\alpha}$ that the left hand side is symmetric in $
\delta
\gamma$, but the right hand side is anti-symmetric. Hence the
Invariant $I^1_{\alpha\beta}$ vanishes if we use the equations of
motion.
\par
There exists Êanother invariant, namely $tr
(M^{-1}\partial_\alpha M J( M^{-1}\partial_\beta M))$ which can be
evaluated to
$$
I^2_{\alpha\beta}\equiv {1\over 2} tr (M^{-1}\partial_\alpha M J(
M^{-1}\partial_\beta M))
=\partial_\alpha x^n\partial_\beta y_n+\nabla_\alpha x_n\nabla_\beta
y^n+(\alpha\leftrightarrow \beta)
$$
$$
=\partial_\alpha x^n\partial_\beta y_n
+\partial_\beta x^n\partial_\alpha y_n
\eqno(2.29)$$
However, setting $\gamma_{\alpha\beta}$ equal to $I^2_{\alpha\beta}$
again leads to a contradiction. In fact there is no manifestly ISO(D,D)
expression for Ê$\gamma_{\alpha\beta}$ that Êleads to a consistent set
of equations of motion.
\par
It might seem that there is Êno manifestly ÊISO(D,D) covariant set of
equations of motion as equations (2.26) and (2.27) do not allow one to
solve for $\gamma_{\alpha\beta}$. The way out is to set
$I^2_{\alpha\beta}=0$ that is adopt in addition to equations (2.26) and
(2.27) the condition
$$
\partial_\alpha x^n\partial_\beta y_n
+\partial_\beta x^n\partial_\alpha y_n=0
\eqno(2.30)$$
Using Êequation (2.26) we can eliminate either $y_n$, or $x^n$, in
equation (2.30) and then find an equation form which we can solve for
$\gamma_{\alpha\beta}$. Eliminating $y_n$ one finds
that
$$
(\sqrt {-\gamma})^{-1}\gamma_{\alpha\beta}=
{\partial_\alpha x^n G_{nm}\partial_\beta x^m\over \sqrt {det
\partial_\alpha x^n G_{nm}\partial_\beta x^m}}
\eqno(2.31)$$
while eliminating $x^n$ Êwe find that
$$
(\sqrt {-\gamma})^{-1}\gamma_{\alpha\beta}=
{\partial_\alpha y_n G^{nm}\partial_\beta y_m\over \sqrt {det
\partial_\alpha y_n G^{nm}\partial_\beta y_m}}
\eqno(2.32)$$
While these are not manifestly ISO(D,D) invariant expressions, they
arise
from equations that are manifestly covariant and Êone can verify that
they
are inert of one uses the equation of motion of equation (2.26).
Substituting these expression for
$(\sqrt {-\gamma})^{-1}\gamma_{\alpha\beta}$ of equation (2.31) into
equation (2.26) and differentiating with respect to $\partial_\alpha$ we
find the standard equation of motion of the bosonic string. On the
otherhand substituting these expression for Ê$(\sqrt
{-\gamma})^{-1}\gamma_{\alpha\beta}$ of equation (2.32) into equation
(2.27) and differentiating with respect to
$\partial_\alpha$ we find the equations of motion in terms of the dual
variable $y_n$.
\par
Hence we have derived from the non-linear realisation a set of
manifestly ÊISO(D,D) covariant equations of motion; for
completeness we summarize them
$$
\epsilon ^{\alpha\beta}J(M^{-1}\partial_\beta M)=\sqrt{-\gamma}
\gamma^{\alpha\beta} M^{-1}\partial_\beta M \ \ \ {\rm or }\ \ \
\epsilon ^{\alpha\beta}\partial_\beta y_n=\sqrt{-\gamma}
\gamma^{\alpha\beta}\nabla_\beta x_n,
$$
$$
tr (M^{-1}\partial_\alpha M J(
M^{-1}\partial_\beta M))=0\ \ Ê\ {\rm or }\ \ \ \partial_\alpha
x^n\partial_\beta y_n +\partial_\beta x^n\partial_\alpha y_n=0
\eqno(2.33)$$

\medskip
{\bf 3. SO(D,D) invariant Hamiltonian formulation of the string}
\medskip
To investigate the quantisation of the SO(D,D) ÊÊstring we
require a Hamiltonian formulation. One might deduce such a formulation
from a Lagrangian that can lead to the string Êmotion
described in term of either Ê$x^\mu$, or the $y_\mu$, depending which
equation of motion one Êchooses to implement first. However, this turns
out to be Êcomplicated involving first and second class constraints
that must be separated. Here we will content ourselves with producing a
Hamiltonian and Poisson brackets that do lead to the SO(D,D) invariant
description of the string motion given in the previous section.
\par
We first introduce the fields
$X^N=(x^\mu, y_\mu)$ which transform according to the vector
representation of SO(D,D). We take as our Hamiltonian
$$
H={1\over 2} \int d\sigma (\delta \partial_1 X^M G_{MN}\partial_1 X^N+
\epsilon \partial_1 X^M \Omega_{MN}\partial_1 X^N )
\eqno(3.1)$$
where $\epsilon$ and $\delta$ are new fields, $G_{MN}$ and
$\Omega^{MN}=\Omega^{NM}$ are given by
$$
G_{MN}=\left(\matrix {g_{\mu\nu}&0\cr 0& (g^{-1})^{\mu\nu}\cr}\right)
,\quad
\Omega^{MN}=\left(\matrix {0&\delta_{\mu}^{\nu}\cr \delta^{\mu}_{\nu}
&0\cr
}\right)
\eqno(3.2)$$
and $\Omega_{MN}=(\Omega^{-1})_{MN}$. Here $g_{\mu\nu}$ is the
background metric. For simplicity we have set the background two form to
zero. ÊThe tensor $\Omega^{MN}$ is an SO(D,D) invariant tensor and one
can verify that $G\Omega G=\Omega^{-1}$. The latter states that $G$
viewed as a matrix is an element of SO(D,D).
\par
We adopt as our Poisson bracket the relation
$$
\{X^N (\sigma ), X^M (\sigma'\}=\theta Ê(\sigma-\sigma')\Omega^{MN}
\eqno(3.3)$$
where $\theta $ is the step function which obeys ${\partial\over
\partial\sigma}
\theta (\sigma-\sigma')= \delta Ê(\sigma-\sigma')$.
\par
Carrying out the Hamiltonian analysis we Êrealise that the momenta
$\rho$ and $\tau$ conjugate to $\delta$ and $\epsilon$ are absent and so
we have the constraints $\rho=0=\tau$. Following the Dirac procedure we
must insist that the time evolution of these constraints vanish that is
$$
\dot \rho=\{\rho, H\}= C_1\equiv\partial_1 X^M G_{MN}\partial_1 X^N=0
\eqno(3.4)$$
and
$$
\dot \tau=\{\tau, H\}=C_2\equiv \partial_1 X^M \Omega_{MN}\partial_1
X^N=0
\eqno(3.5)$$
Thus we find two new constraints which obey the Poisson brackets
$$
\{C_1(\sigma),C_1 (\sigma')\}= -4 C_1 (\sigma')\partial_1 \delta
(\sigma-\sigma')+2\partial_1 C_1(\sigma) \delta (\sigma-\sigma'),
$$
$$
\{C_2(\sigma),C_2 (\sigma')\}= -4 C_2 (\sigma')\partial_1 \delta
(\sigma-\sigma')+2\partial_1 C_2(\sigma) \delta (\sigma-\sigma'),
$$
$$
\{C_1(\sigma),C_2 (\sigma')\}= -4 C_1 (\sigma')\partial_1 \delta
(\sigma-\sigma')+2\partial_1 C_1(\sigma) \delta (\sigma-\sigma')
\eqno(3.6)$$

These constraints Êare first class and generate the
Virasoro algebra as expected. We note that this also ensures that taking
the time evolution of the constraints of equation (3.4) and (3.5)
generates no new constraints.
\par
The equation of motion of $X^M$ is given by
$$
\partial_0 X^M =\{X^M, H\}=\delta \Omega ^{MN}G_{NP}\partial_1 X^P+
\epsilon \partial_1 X^M
\eqno(3.7)$$
which we may write in matrix form as
$$
\partial_0 X =(\epsilon +\delta \Omega G)\partial_1 X
\eqno(3.8)$$
This in turn implies that
$$
\partial_1 X ={1\over \delta^2-\epsilon^2}(-\epsilon +\delta \Omega
G)\partial_0 X
\eqno(3.9)$$
Introducing the two new variable $\tilde \gamma ^{\alpha\beta}=\sqrt
{-\gamma }\gamma^{\alpha\beta}$, where $\gamma =\det
\gamma_{\alpha\beta}$ by setting
$$
\delta =\tilde \gamma_{00},\quad \epsilon= -{\tilde \gamma^{01}\over
\tilde \gamma^{00}}
\eqno(3.10)$$
and substituting into equations (3.8) and (3.9) we find that they are
the
same as
$$
\Omega_{MN}\epsilon^{\alpha\beta}\partial_\beta X^N= \sqrt {-\gamma
}\gamma^{\alpha\beta}G_{MN} \partial_\beta X^N
\eqno(3.11)$$
For example, taking $\alpha=0$ in this latter equation, bring all the
$\partial_1X$ terms to one side and dividing to by $\tilde \gamma^{00}$
we find equation (3.8).
\par
We next show that the constraints of equations (3.4) and (3.5) Êare
equivalent to the condition
$$
\partial_\alpha X^M \Omega_{MN}\partial _\beta X^N=0
\eqno(3.12)$$
provided one uses the equation of motion of $X^M$. Taking $\alpha=1=
\beta$
we find the constraint of equation (3.5). Substituting the
equation of motion of equation (3.11) with $\alpha=0$ that is the
equation
$$
\partial_1 X= \delta^{-1} \Omega G \partial_0 X-\epsilon \Omega G
\partial_1 X
\eqno(3.13)$$
we find, using equation (3.5), that
$$
\partial_1 X^M G_{MN}\partial _1 X^N=0
\eqno(3.14)$$
Proceeding in this way we find the constraint of equation (2.30), or
equivalently equation (2.33), Êas well as Êthe constraint
$$
\partial_\alpha X^M G_{MN}\partial _\beta X^N=0
\eqno(3.15)$$
In fact this is not an independent Êconstraint as it follows from
that of
equation (3.5) using the equation of motion of equation (3.11).
\par
Thus the Hamiltonian system introduced above is equivalent to the
motion of
equation (2.33) and so we can be confident that it is the correct
Hamiltonian system.
%%%%%%%%%%%%%%%%%%%%%%%%%%%%%%%%%%%%%%%%%%%%%%%%%%%%%%%%%%
\medskip
{\bf 4. Quantisation of the SO(D,D) symmetric string }
\medskip
To quantise the SO(D,D) formulation of the string is straightforward.
The Poisson brackets of equation (3.3) become the commutators
$$
[\hat ÊX^M (\sigma ),\hat X^N (\sigma'] =i\Omega^{MN}\theta
(\sigma-\sigma')
\quad {\rm or \ equivalently } \quad [\hat x^\mu (\sigma ),\hat y_\nu
(\sigma' ) ]= i\delta^\mu_\nu \theta Ê(\sigma-\sigma')
\eqno(4.1)$$
The Hamiltonian of equation (3.1) and constraints of equations
(3.4) and (3.5) now contain operator valued
$X^M$'s. We impose the constraints on the wavefunction
$$
\hat C_1 \Psi=0=\hat C_2 \Psi
\eqno(4.2)$$
The ÊSchr\"odinger equation then states that Êthe
wavefunction is independent time as the Hamiltonian on it now vanishes.
In fact such a commutator was suggested in [6] on the grounds Êthat
the
$y_\mu$ coordinates are related to the momenta of the theory with just
$x^\mu$ in the linearised theory.
\par
Following the same Êtreatment used to derive the standard
uncertainty principle we find that
$$
(\Delta x^\mu )^2 (\Delta y_\nu )^2= <x|(x^\mu-<x^\mu>)^2 |x>
<y|(y_\nu-<y_\nu>)^2 |y>
$$
$$
\ge
|<x|(x^\mu-<x^\mu>)(y_\nu-<y_\nu>) |y>|^2
\ge \delta^\mu_\nu \theta Ê(\sigma-\sigma')|<x|y>|^2
\eqno(4.3)$$
Hence, one can not measure both $x^\mu$ and $y^\nu$. This is
consistent with the well known observation that when Êa string is
wrapped on a circle only distances down to Êa minimum radius are
observable.
\par
The simplest way to proceed is to choose the operators to be given by
$$
\hat x^\mu (\sigma)=x^\mu (\sigma)\quad {\rm and} \quad \hat y_\mu
(\sigma)= -i\int ^\sigma d\sigma' {\delta \over \delta x^\mu (\sigma')}
\eqno(4.4)$$
In this case the wavefunction depends on $x^\mu (\sigma)$ and we
arrive at the standard picture of the second quantised bosonic
string as studied in [17].
\par
However, we could equally well take the representation
$$
\hat x^\mu (\sigma)=i\int ^\sigma d\sigma' {\delta \over \delta y_\mu
(\sigma')} \quad {\rm and} \quad \hat y_\mu (\sigma)= Êy_\mu (\sigma)
\eqno(4.5)$$
\par
The relation between the two representations is given by
$$\psi [ x^\mu (\sigma)]= < x^\mu (\sigma)|\psi>= \int {\cal D} y_\nu
(\sigma ') e^{i\int d\sigma" x^\nu(\sigma")\partial_1 y_\nu (\sigma")}
\psi [y(\sigma)]
\eqno(4.6)$$
\par
Thus although one can use either representation, or any representation
that is related by a SO(D,D) rotations, by choosing a
representation Êone makes a choice and breaks the manifest SO(D,D)
symmetry. However, this is not an actually breaking of this symmetry as
considering all representations on an equal footing preserves SO
(D,D). ÊHowever, to
work in a way that Êmanifestly preserves the SO(D,D) symmetry one must
keep both $x^\mu$ and $y_\mu$ and use the techniques of ÊÊnon-
commutative
field theory.
\par
The $E_{11}\otimes_s l_1$ non-linear realisation, viewed from
the IIA perspective, and at ÊÊlowest level is just the ÊSO(D, D) string.
Hence quantising Êthe string dynamics that follows from this non-linear
realisation leads to the same results at lowest order. In general we
expect the higher level effects to follow the same pattern; the
coordinates will obey non-trivial commutation relations and in order to
keep the symmetry manifest one must work with a non-commutative field
theory. It is likely that in general the set of commuting coordinates is
larger than the set of space-time generators; for example in the
dimensions lower than ten we would expect the generators of spacetime
translations associated with a torus Êdimensional reduction to commute.
\par
The adoption of just the space-time translations in the non-linear
realisation of
$E_{11}$ has, in a number of circumstances, Êworked better than one
might
expect given that the next coordinate is only one level more than the
usual translations. The analysis given here suggests Êthat keeping only
the space-time translations ÊÊpreserves more of the
$E_{11}$ symmetry than expected. It would certainly be good to
understand how much of the symmetry is hidden in this way and how
much is
automatically encoded by the existence of Êdifferent representations.
\par
The string resulting from the choice of coordinates of the
representation
given in Êequation (4.4) contains as massless fields the graviton,
antisymmetric tensor field and tachyon which depend only on $x^\mu$.
Apart from the tachyon these are the fields that are contained in a
non-linear realisation
of ISO(D,D) with local subgroup $SO(D) \otimes SO(D)$. However, for the
choice of equation (4.5) we find the same fields but they now Êdepend on
$y_\mu$. These two formulations are related by an SO(D,D) rotation.
However, a manifestly ÊSO(D,D) invariant formulation requires both
$x^\mu$ and
$y_\mu$ but, as we have just pointed out, Êit is not a usual quantum
fields theory but Êa non-commutative field theory.
\par
It would be interesting to repeat this calculations given in this paper
for the membrane in eleven dimensions where the lowest level coordinates
are
$x^a$ and $x_{ab}$. These are likely to obey non-trivial commutation
relations whose right hand sides are field dependent.

\medskip
{\bf References}
\medskip
\item{[1]} P. West, {\sl $E_{11}$ and M Theory}, Class. Quant.
Grav. {\bf 18 } (2001) 4443, {\tt hep-th/0104081}
\item{[2]} P.~C. West, {\sl Hidden superconformal symmetry in {M}
ÊÊÊÊtheory }, ÊJHEP {\bf 08} (2000) 007, {\tt hep-th/0005270}
\item{[3]} P. West, {\sl $E_{11}$, SL(32) and Central Charges},
Phys. Lett. {\bf B 575} (2003) 333-342, {\tt hep-th/0307098}
\item{[4]} P. West, {\sl $E_{11}$ origin of brane charges and U-duality
ÊÊmultiplets}, {\bf ÊJHEP} 0408 (2004) 052, hep-th/0406150.
\item{[5]} M. Duff, ``Duality Rotations In String Theory,''
ÊÊNucl.\ Phys.\ ÊB {\bf 335} (1990) 610; M. Duff and J. Lu,
{\sl Duality rotations in
membrane theory}, ÊNucl. Phys. {\bf B347} (1990) 394.
\item{[6]} Tseytlin, Phys.Lett. {\bf B242} (1990) 163.
\item{[7]} A. Kleinschmidt and ÊP. West, {\sl
Representations of
${\cal G}^{+++}$ and the role of space-time}, JHEP {\bf 0402} (2004)
033,
hep-th/0312247.
\item{[8]} P. West, Brane dynamics, central charges and $E_{11}$,
JHEP 0503 (2005)
077, hep-th/0412336.
\item{[9]} P. Cook and P. West, Charge multiplets and masses for E(11);
JHEP 11 (2008) 091, ÊarXiv:0805.4451.
\item{[10]}
S. Elitzur, A. Giveon, D. Kutasov and E. ÊRabinovici, Ê{\it
Algebraic aspects of matrix theory on $T^d$ }, {hep-th/9707217};
\item{[11]} ÊN. Obers, ÊB. Pioline and E.
Rabinovici, {\it M-theory and U-duality on $T^d$ with gauge
backgrounds},
{ hep-th/9712084};
\item{[12]} ÊN. Obers and B. Pioline,~ {\it U-duality and
M-theory, an
algebraic approach}~, { hep-th/9812139};
\item{[13]} ÊN. Obers and B. Pioline,~ {\it U-duality and
M-theory}, {hep-th/9809039}.
\item {[14]} F. Riccioni and P. West, E(11)-extended spacetime and
gauged
supergravities, JHEP0802:039,2008; ÊarXiv:0712.1795
\item {[15]} P. West $E_{11}$ and Higher Spin Theories,
Phys.Lett. {\bf B650}, 197,2007, Êhep-th/0701026.
\item{[16]} P. ÊGoddard, J. Goldstone, C. ÊRabbi and
C. Thorn; Nucl. Phys. {\underbar {B56}}, 109 (1973).
\item{[17]} S. Coleman, J. Wess and ÊB. Zumino, ÊÊ{\sl Structure of
Phenomenological Lagrangians. 1}, Phys.Rev. {\bf 177} (1969) 2239;
C. Callan, S. Coleman, J. Wess and B. Zumino, Phys.
Rev. 177 (1969) 2239; 2247.

\medskip
$$
\matrix{
& & & & &&& &\bullet &11&&&
\cr & & & &&& & &| & && &
\cr
\bullet&-&\bullet&-&\ldots &- &\bullet&-&\bullet&-&\bullet&-&\bullet
\cr
1& &2& & & &7& &8& & 9&
&10\cr}
$$
\par
\centerline {Fig 1. The $E_{11}$ Dynkin diagram}

\medskip

%bars if any missing on top 2 at end Êand last of seven dim.
$$\halign{\centerline{#} \cr
\vbox{\offinterlineskip
\halign{\strut \vrule \quad \hfil # \hfil\quad &\vrule Ê\quad \hfil #
\hfil\quad &\vrule \hfil # \hfil
&\vrule \hfil # \hfil Ê&\vrule \hfil # \hfil &\vrule \hfil # \hfil &
\vrule \hfil # \hfil &\vrule \hfil # \hfil &\vrule \hfil # \hfil &
\vrule \hfil # \hfil &\vrule#
\cr
\noalign{\hrule}
D&G&$Z$&$Z^{a}$&$Z^{a_1a_2}$&$Z^{a_1\ldots a_{3}}$&$Z^{a_1\ldots a_
{4}}$&$Z^{a_1\ldots a_{5}}$&$Z^{a_1\ldots a_6}$&$Z^{a_1\ldots a_7}$&\cr
\noalign{\hrule}
8&$SL(3)\otimes SL(2)$&$\bf (3,2)$&$\bf (\bar 3,1)$&$\bf (1,2)$&$\bf
(3,1)$&$\bf (\bar 3,2)$&$\bf (1,3)$&$\bf (3,2)$&$\bf (6,1)$&\cr
&&&&&&&$\bf (8,1)$&$\bf (6,2)$&$\bf (18,1)$&\cr Ê&&&&&&&$\bf (1,1)$&&$
\bf
(3,1)$&\cr Ê&&&&&&&&&$\bf (6,1)$&\cr
&&&&&&&&&$\bf (3,3)$&\cr
\noalign{\hrule}
7&$SL(5)$&$\bf 10$&$\bf\bar 5$&$\bf 5$&$\bf \bar {10}$&$\bf 24$&$\bf
40$&$\bf 70$&-&\cr Ê&&&&&&$\bf 1$&$\bf 15$&$\bf 50$&-&\cr
&&&&&&&$\bf 10$&$\bf 45$&-&\cr
&&&&&&&&$\bf 5$&-&\cr
\noalign{\hrule}
6&$SO(5,5)$&$\bf \bar {16}$&$\bf 10$&$\bf 16$&$\bf 45$&$\bf \bar
{144}$&$\bf 320$&-&-&\cr &&&&&$\bf 1$&$\bf 16$&$\bf 126$&-&-&\cr
&&&&&&&$\bf 120$&-&-&\cr
\noalign{\hrule}
5&$E_6$&$\bf\bar { 27}$&$\bf 27$&$\bf 78$&$\bf \bar {351}$&$\bf
1728$&-&-&-&\cr Ê&&&&$\bf 1$&$\bf \bar {27}$&$\bf 351$&-&-&-&\cr
&&&&&&$\bf 27$&-&-&-&\cr
\noalign{\hrule}
4&$E_7$&$\bf 56$&$\bf 133$&$\bf 912$&$\bf 8645$&-&-&-&-&\cr
&&&$\bf 1$&$\bf 56$&$\bf 1539$&-&-&-&-&\cr
&&&&&$\bf 133$&-&-&-&-&\cr
&&&&&$\bf 1$&-&-&-&-&\cr
\noalign{\hrule}
3&$E_8$&$\bf 248$&$\bf 3875$&$\bf 147250$&-&-&-&-&-&\cr
&&$\bf1$&$\bf248$&$\bf 30380$&-&-&-&-&-&\cr
&&&$\bf 1$&$\bf 3875$&-&-&-&-&-&\cr
&&&&$\bf 248$&-&-&-&-&-&\cr
&&&&$\bf 1$&-&-&-&-&-&\cr
\noalign{\hrule}
}}\cr}$$
\par

\centerline {Table 1. The Brane ÊCharge representations of
the group, G, derived from the $l_1$ representation of $E_{11}$}

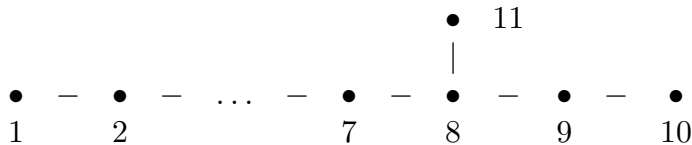
\end